\documentclass[useAMS,usenatbib]{mn2e}
\usepackage{amssymb}
\usepackage{graphicx}

\title[]{The dispersion of period spacing for DAV stars}
\author[Y. H. Chen]{Y. H. Chen\thanks{E-mail: yhc1987@cxtc.edu.cn}\\
School of Physics and Electronical Science, Chuxiong Normal University, Chuxiong 675000, China}

\begin{document}

\date{Accepted: }

\pagerange{\pageref{firstpage}--\pageref{lastpage}} \pubyear{????}

\maketitle

\label{firstpage}

\begin{abstract}

Three groups of DAV star models are evolved with time-dependent element diffusion by \texttt{WDEC}. The core compositions of these models are directly from white dwarf models evolved by \texttt{MESA}, which are results of thermonuclear burning. Based on these DAV star models, we try to study the dispersion of period spacing. The thickness of hydrogen atmosphere can seriously affect the deviation degree of minimal period spacings. The minimal period spacings dominate the dispersion of period spacing. The thinner the hydrogen atmosphere, basically, the more dispersive the period spacing. Standard deviations are used to analyze the dispersion of period spacing. Studying the dispersion of period spacing on a DAV star KUV03442+0719 preliminarily, we suggest that log($M_{\rm H}/M_{*}$) is from -8.5 to -5.5. In addition, modes partly trapped in C/O core are found based on those DAV star models. The identified modes and average period spacings indicate that KUV03442+0719 may be the first star to 'observe' modes partly trapped in C/O core.

\end{abstract}

\begin{keywords}
asteroseismology-stars: individual(KUV03442+0719)-white dwarfs
\end{keywords}

\section{Introduction}

Asteroseismology is an unique and powerful tool to detect the inner structure of stars. About 98\% of all stars will evolve to be white dwarfs ($WD$s) and about 80\% of them show DA spectral class (Fontaine, Brassard, \& Bergeron 2001, Winget, \& Kepler 2008, Bischoff-Kim, \& Metcalfe 2011). A DA type $WD$ is composed of a compact and degenerate carbon/oxygen (C/O) core and a very thin helium (He) layer covered by a thinner hydrogen (H) atmosphere. At the bottom of the $WD$ cooling curve, there is an instability stripe basically from 12270\,K to 10850\,K (Gianninas et al. 2005, Gianninas et al. 2006) for DAV stars. At least 148 DAV stars were now observed (Castanheira et al. 2010). Doing asteroseismological studies on 83 DAV stars, Castanheira, \& Kepler (2009) found an average surface H atmosphere of $M_{\rm H}/M_{*}$ = $10^{-6.3}$ and a range of $M_{\rm H}/M_{*}$ from $10^{-9.5}$ to $10^{-4.0}$. The average surface H atmosphere is thinner than previous predicted value of $M_{\rm H}/M_{*}$ = $10^{-4.0}$. Romero et al. (2012) did asteroseismological studies on 44 bright DAV stars based on fully evolutionary DA type $WD$ models. The detailed chemical profiles from center to surface were concerned. Measuring the changing rate of main pulsation period of G117-B15A, Kepler et al. (2000) constrained the evolutionary timescale of the star. The changing rate of the pulsation mode is (2.3 $\pm$ 1.4) $\times$ $10^{-15}$ s $s^{-1}$, which makes G117-B15A the most stable optical clock known. Massive DAV stars can be used to test the crystallization theory (Montgomery, \& Winget 1999). Based on an analysis of the average period spacing of BPM 37093, Kanaan et al. (2005) suggested that a large fraction of BPM 37093 was likely to be crystallized. Above all, the asteroseismological studies on DAV stars are significant and meaningful.

DAV stars are $g$-mode pulsators. Gravity acts as restoring force. An eigen-mode can be expressed by three indices ($k$, $l$, $m$). They are the radial order, the spherical harmonic degree, and the azimuthal number respectively. The radial order $k$ describes the nodes of standing wave inside a star, which can not be identified observationally. According to an asymptotic theory for $g$-modes (Tassoul 1980), modes with same $l$ and consecutive $k$ (high values) have an asymptotic period spacing as
\begin{equation}
\bar{\triangle \texttt{$P$}(l)}=\frac{2\pi^{2}}{\sqrt{l(l+1)}{\int_{0}}^{R}\frac{|N|}{r}dr}.
\end{equation}
\noindent In Eq.\,(1), $N$ is Brunt-V\"ais\"al\"a frequency and $R$ is stellar radius. The Eq.\,(1) is valid when radial orders are sufficiently high. If mode trapping effects are negligible and ($l$, $m$) values can be identified, the relative radial orders for $m$ = 0 modes with same $l$ can be counted under the linear approximations.

The spherical harmonic degree $l$ of some mode can be identified according to a frequency splitting relation. A pulsating mode will split into several components in frequency because of stellar rotation. Brickhill (1975) reported an approximate formula between frequency splitting ($\delta\nu_{k,l}$) and rotational period
($P_{\rm rot}$) under the asymptotic regime as
\begin{equation}
m\delta\nu_{k,l}=\nu_{k,l,m}-\nu_{k,l,0}=\frac{m}{P_{\rm rot}}(1-\frac{1}{l(l+1)}).
\end{equation}
\noindent In Eq.\,(2), the azimuthal number $m$ can be taken from $-l$ to $l$, totally 2$l$+1 different values. Modes with $l$ = 1 will form a triplet and modes with $l$ = 2 will form a quintuplet. Therefore, if triplets or quintuplets are derived from observations for a pulsating star, the spherical harmonic degree $l$ will be identified as 1 or 2 preferentially to some extent. Sometimes, three middle components of quintuplets are detected, which will confuse the $l$ identifications.

Normal $g$-modes are global to the star which pulsate within its entire body. Winget et al. (1981) proposed that compositional stratified zone rapidly produced by gravity sedimentation may form a resonance. Some modes would be trapped in the resonance. They were called trapped modes. Brassard et al. (1992a) calculated the period spacing between trapped modes ($\Pi_{\rm H}$ in Brassard et al. (1992a)) and the period spacing between normal modes ($\Pi_{\rm 0}$ in Brassard et al. (1992a)). Trapped modes and normal modes have different period spacings. Brassard et al. (1992a) reported that the thinner the H atmosphere, the larger the period spacing for modes trapped in H layer. On the period ($P$($k$)) versus period spacing ($P$($k$+1)-$P$($k$)) diagram, trapped modes show minimal period spacings. The five identified trapped modes ($l$ = 1) for the DOV star PG1159-035 show minimal period spacings on the period versus period spacing diagram (Costa et al. 2008). Brassard et al. (1992a) made a detailed study on mode trapping properties for DAV stars. They showed that trapped modes had minimal values on the period versus period spacing diagram. In addition, they reported that the thinner the H atmosphere, the more obvious the mode trapping property. C$\acute{o}$rsico et al. (2002) also discussed mode trapping properties at length on a 0.563\,$M_{\odot}$ DAV star evolved by \texttt{LPCODE}. \texttt{LPCODE} is a program to evolve $WD$s, which made full evolutionary $WD$ models taking time-dependent element diffusion into account. They reported, as expected in the context of the trace element approximation of equilibrium profiles (Tassoul et al. 1990), that mode trapping properties were stronger when compared with composition profiles obtained at, or near, diffusion equilibrium.

DAV stars have low luminosity which is basically less than 1\% of that for the sun ($LogL/L_{\odot}$ $<$ -2). Usually only a few modes are observed for DAV stars. It is difficult to find trapped modes on the period versus period spacing diagram. Since a trapped mode brings a minimal period spacing, it should cause a large dispersion of period spacing. Therefore, we try to study the dispersion of period spacing for DAV stars. In Sect. 2, input physics and model calculations are discussed. In Sect. 3, we try to study the dispersion of period spacing for DAV stars. The effect of different thicknesses of H atmosphere is discussed in Sect. 3.1. The effect of different values of effective temperature and He layer mass is discussed in Sect. 3.2. The effect of different values of total stellar mass are discussed in Sect. 3.3. The average period spacing for DAV star models is analyzed in Sect. 4. In Sect. 5, we preliminarily study the dispersion of period spacing for a DAV star KUV03442+0719. In Sect. 6, discussion and conclusions are summarized.

\section{Input physics and model calculations}

In this Section, we discuss the input physics and model calculations. In 2011, Paxton et al. reported modules for experiments in stellar astrophysics (\texttt{MESA}) which can evolve a star from main sequence ($MS$) stage to $WD$ stage. \texttt{MESA} is updated quickly on the web site of 'http://mesa.sourceforge.net/'. The core compositions of a $WD$ star are results of thermonuclear burning. The thermonuclear reaction rates are from Caughlan \& Fowler (1988) and Angulo et al. (1999) for \texttt{MESA}. With \texttt{MESA} version 4298, we evolve a group of MS stars, as shown in the first column in Table 1, to $WD$ stage by a module named make\_co\_wd. The input settings in the module are default by \texttt{MESA}. When these progenitor stars evolve to the $WD$ cooling curve on the Hertzsprung-Russell ($H-R$) diagram, the $WD$ models are took out. The core composition structures of these $WD$ models, including mass, radius, luminosity, pressure, temperature, entropy, and carbon profile, are took out together. The location of maximal carbon profile is used as the boundary of the core for approximate calculations. The evolved $WD$ masses in \texttt{MESA} and their core masses are showed in Table 1. Using the module make\_co\_wd to evolve $WD$s is quick but a little rough. In the future work, we will try to evolve $WD$s in the module named 1M\_pre\_ms\_to\_wd and refine the input settings.

\begin{table}
\begin{center}
\begin{tabular}{lllll}
\hline
$MS$         &$WD(\texttt{MESA})$   &$M_{core}(\texttt{MESA})$ &$WD(\texttt{WDEC})$     \\
\hline
($M_{\odot}$)&($M_{\odot}$)         &($M_{\odot}$)             &($M_{\odot}$)           \\
\hline
2.0          &0.580                 &0.555                     &0.550-0.570             \\
2.8          &0.614                 &0.595                     &0.580-0.600             \\
3.0          &0.633                 &0.615                     &0.610-0.630             \\
3.2          &0.659                 &0.645                     &0.640-0.660             \\
3.4          &0.689                 &0.675                     &0.670-0.690             \\
3.6          &0.723                 &0.715                     &0.700-0.720             \\
3.8          &0.751                 &0.740                     &0.730-0.750             \\
4.0          &0.782                 &0.770                     &0.760-0.780             \\
4.5          &0.805                 &0.800                     &0.790-0.810             \\
5.0          &0.832                 &0.825                     &0.820-0.830             \\
\hline
\end{tabular}
\caption{Masses of $MS$ stars, $WD$ stars evolved by \texttt{MESA}, corresponding cores, and corresponding $WD$ stars evolved by \texttt{WDEC}.}
\end{center}
\end{table}

The structure parameters are inserted into white dwarf evolution code (\texttt{WDEC}). \texttt{WDEC} was first written by Schwarzschild and subsequently modified by Kutter \& Savedoff (1969) (\texttt{WDEC}1.0), Lamb \& van Horn (1975) (\texttt{WDEC}2.0), Winget (\texttt{WDEC}3.0), Kawaler (\texttt{WDEC}4.0), Wood 1990 (\texttt{WDEC}5.0), and Bradley (\texttt{WDEC}6.0). The program is from White Dwarf Research Corporation (http://whitedwarf.org/). \texttt{WDEC} can evolve groups of DAV star models. The model parameters are adopted as total stellar mass, effective temperature, He layer mass fraction, and H atmosphere mass fraction. With the core composition structures from $WD$ models evolved by \texttt{MESA}, groups of $WD$ models are evolved by \texttt{WDEC}. The He layer and the H atmosphere are very thin for a DAV star. Therefore, we take the total stellar mass of $WD$ models in \texttt{WDEC} the same as the $WD$ core masses in \texttt{MESA}, as shown in Table 1. The carbon profiles of those $WD$ models are results of thermonuclear burning.

The element diffusion scheme of Thoul et al. (1994) are added into \texttt{WDEC} by Su et al. (2014a). The Eq.\,(40) of Thoul et al. (1994) is used to solve the problem. The Eq.\,(21) of Thoul et al. (1994) is used to calculate diffusion coefficients. The usual Coulomb potential is used and the cross section for Coulomb scattering are from Eq.\,(8) and Eq.\,(9) of Thoul et al. (1994). The pressure gradient, the temperature gradient, and the concentration gradient are included, as shown in Eq.\,(41) of Thoul et al. (1994). The ideal gas equation of state is used and the effect of degeneracy is not included. Therefore, there may be some problems in the application of C/O core. We do not use an ordinary differential equation solver package, but instead of an implicit differential scheme described by Zhang (2008). The H/He and He/C interfaces are time-dependent element diffusion results. Paquette, Pelletier, Fontaine, \& Michaud (1986a) reported that the use of a pure Coulomb potential on $WD$s can rise serious errors in the diffusion coefficients even in the atmospheres and the envelopes. In order to obtain much more reliable diffusion coefficients, the screened Coulomb potential should be used in our future work (Paquette, Pelletier, Fontaine, \& Michaud 1986a, Paquette, Pelletier, Fontaine, \& Michaud 1986b, Cox, Guzik, \& Kidman 1989). According to the work of Bergeron et al. (1995), the mixing length parameter is adopted as 0.6. For more details about the input physics please refer to our previous paper studying EC14012-1446 (Chen \& Li 2014).

The total stellar mass of DAV stars is basically around 0.6\,$M_{\odot}$ (Gianninas, Bergeron, \& Ruiz 2011). We evolve three groups of $WD$ models. For the first group, the total stellar mass ($M_{*}$) is 0.600\,$M_{\odot}$ and the effective temperature ($T_{\rm eff}$) is 12000\,K. The He layer mass fraction (log($M_{\rm He}/M_{*}$)) is from -4.0 to -2.0 with steps of 0.5 and the H atmosphere mass fraction (log($M_{\rm H}/M_{*}$)) is from -10.0 to -4.0 with steps of 0.5. For the second group, $M_{*}$ is 0.600\,$M_{\odot}$, $T_{\rm eff}$ is from 12300\,K to 10800\,K with steps of 50\,K, log($M_{\rm He}/M_{*}$) is -2.0, and log($M_{\rm H}/M_{*}$) is from -10.0 to -4.0 with steps of 0.5. For the third group, $M_{*}$ is from 0.550\,$M_{\odot}$ to 0.830\,$M_{\odot}$ with steps of 0.010\,$M_{\odot}$, $T_{\rm eff}$ is 12000\,K, log($M_{\rm He}/M_{*}$) is -2.0, and log($M_{\rm H}/M_{*}$) is from -10.0 to -4.0 with steps of 0.5. More than 800 $WD$ models are evolved. We use a modified pulsation code of Li (1992a,b) to calculate the pulsation frequencies. The reflective boundary condition, namely Lagrangian pressure perturbation vanished at stellar radius, is adopted. The square of Brunt-V\"ais\"al\"a frequency is calculated from Eq.\,(14), not Eq.\,(1) of Brassard et al. (1991). In the degenerate C/O core, the square of Brunt-V\"ais\"al\"a frequency is close to zero. The full equations of linear and adiabatic oscillation are solved numerically and then the eigenfrequencies are scanned successively. As talked above, the $WD$s evolved in the module of make\_co\_wd is relatively rough. It was mentioning that the corresponding core compositions may affect the Brunt-V\"ais\"al\"a frequency and then affect the eigenfrequency. Tassoul, Fontaine, \& Winget (1990) proposed that the accuracy of $l$ = 1, 2, and 3 modes would increase more than 1\% if the shell number was made dense from 160-190 to 600. The shell numbers of cores from $WD$ models evolved by \texttt{MESA} are around 260. For the $WD$ models evolved by \texttt{WDEC}, the shell numbers are around 500 and the corresponding shell numbers of their cores are around 170. The core information with 260 shells in \texttt{MESA} are detailed enough to make a core with 170 shells in \texttt{WDEC}. We try to do a preliminary analysis on the dispersion of period spacing. We do not make dense the shell number of $WD$ models. The reciprocal of eigenfrequency is eigenperiod. With calculated eigenperiods of those DAV star models, the dispersion of period spacing is studied.

\section{The dispersion of period spacing for DAV stars}

According to the asymptotic theory for $g$-modes, modes with same $l$ and consecutive $k$ (high values) have an asymptotic period spacing, as shown in Eq.\,(1). We first study the $l$ = 1 and 2 modes in a fixed period range from 800 seconds (s) to 1600 seconds (s). In the fixed range, the dispersion of period spacing are calculated quantitatively.

In order to describe the dispersion of period spacing quantitatively, an average period spacing ($APS$) is first introduced by
\begin{equation}
\ APS = \frac{1}{n}\sum_{n}(P(k+1)-P(k)).
\end{equation}
\noindent In Eq.\,(3), $n$ is the number of $P(k)$ locating in the range from 800\,s to 1600\,s. The sum of period spacing is divided by the number of period spacing, which is named as $APS$. In fact, there are $n$ periods in the range and $n$ period spacings calculated. In the range from 800 s to 1600 s, the period spacing of the maximal mode equals the first mode larger than 1600 s (out of the range) minus the maximal mode in the range. The standard deviation of the dispersion of period spacing ($\sigma_{SD}$) is described by
\begin{equation}
\sigma_{SD}=\sqrt{\frac{1}{n} \sum_{n}(P(k+1)-P(k)-APS)^2}.
\end{equation}
\noindent In Eq.\,(4), there are $n$ modes located in the period range from 800\,s to 1600\,s and $n$ period spacings calculated. The parameter $\sigma_{SD}$ is used to study the dispersion of period spacing. The larger the parameter, the more dispersive the period spacing.

\subsection{The effect of different thicknesses of H atmosphere}

Brassard et al. (1992) reported that the thinner the H atmosphere, the more obvious the mode trapping property. Trapped modes make minimal period spacings. Those minimal period spacings are relatively far from the average period spacing. Therefore trapped modes will make large dispersions of period spacing. We first study the effect of different thicknesses of H atmosphere on the dispersion of period spacing. The first group of models are used. For the He layer mass fraction, log($M_{\rm He}/M_{*}$) is adopted as -2.0. The thickness of H atmosphere changes with the other parameters unchangeable.

\begin{figure}
\begin{center}
\includegraphics[width=8cm,angle=0]{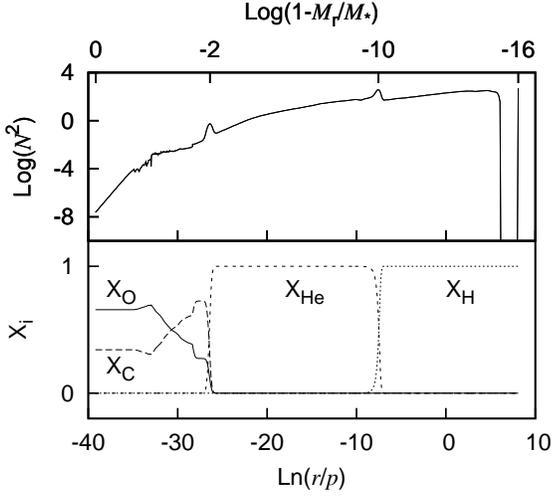}
\end{center}
\caption{The core composition profiles and corresponding Brunt-V\"ais\"al\"a frequency for the model of $M_{*}$ = 0.600\,$M_{\odot}$, $T_{\rm eff}$ = 12000\,K, log($M_{\rm He}/M_{*}$) = -2.0, and log($M_{\rm H}/M_{*}$) = -10.0.}
\end{figure}

\begin{figure}
\begin{center}
\includegraphics[width=8cm,angle=0]{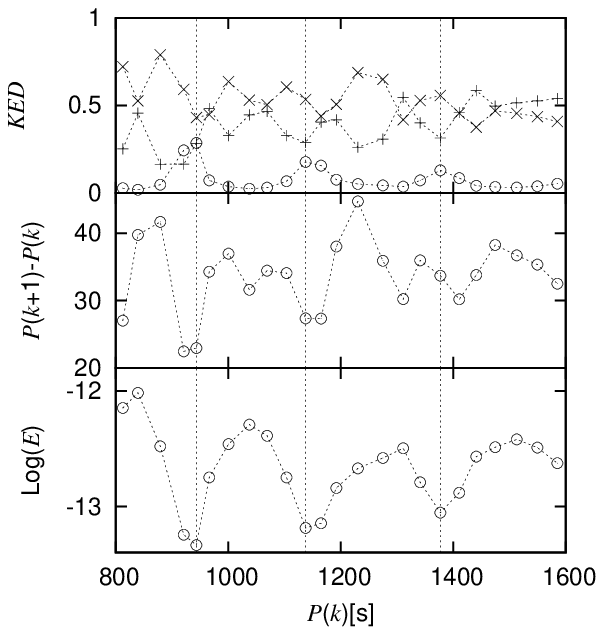}
\end{center}
\caption{Diagram of period versus mode inertia, period versus period spacing, and kinetic energy distribution ($KED$). The model has H atmosphere of log($M_{\rm H}/M_{*}$) = -10.0. The parameter $l$ equals 2 and $k$ is from 22 to 45. In the up panel, open dots, forks, and pluses represent the scale of kinetic energy distributed in H atmosphere, He layer, and C/O core respectively.}
\end{figure}

\begin{figure}
\begin{center}
\includegraphics[width=8cm,angle=0]{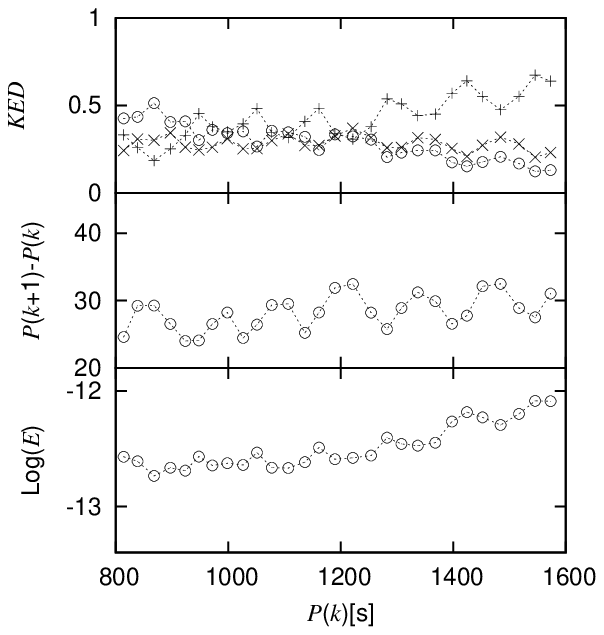}
\end{center}
\caption{Diagram of period versus mode inertia, period versus period spacing, and kinetic energy distribution ($KED$). The model has H atmosphere of log($M_{\rm H}/M_{*}$) = -4.0. The parameter $l$ equals 2 and $k$ is from 29 to 56. In the up panel, open dots, forks, and pluses represent the scale of kinetic energy distributed in H atmosphere, He layer, and C/O core respectively.}
\end{figure}

\begin{figure}
\begin{center}
\includegraphics[width=8cm,angle=0]{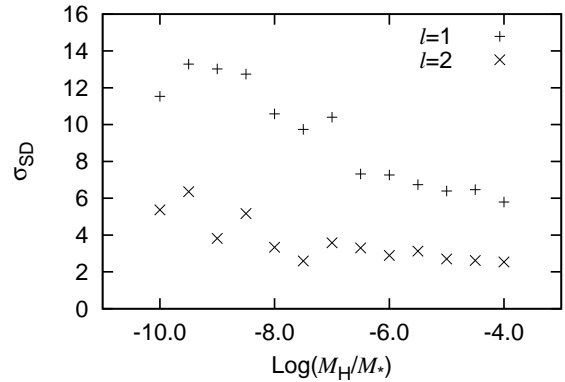}
\end{center}
\caption{Diagram of standard deviation of the dispersion of period spacing ($\sigma_{SD}$). The model parameters are $M_{*}$ = 0.600\,$M_{\odot}$, $T_{\rm eff}$ = 12000\,K, log($M_{\rm He}/M_{*}$) = -2.0. The H atmosphere log($M_{\rm H}/M_{*}$) is from -10.0 to -4.0 with steps of 0.5. The values of $\sigma_{SD}$ are showed in pluses for $l$ = 1 modes and in forks for $l$ = 2 modes.}
\end{figure}

In Fig. 1, we show the core composition profile and Brunt-V\"ais\"al\"a frequency for the model of $M_{*}$ = 0.600\,$M_{\odot}$, $T_{\rm eff}$ = 12000\,K, log($M_{\rm He}/M_{*}$) = -2.0, and log($M_{\rm H}/M_{*}$) = -10.0. The $WD$ model is evolved by \texttt{WDEC} with time-dependent element diffusion. The carbon profile of the core is from an evolution of 2.8\,$M_{\odot}$ $MS$ progenitor star by \texttt{MESA}. We can see that the H/He and He/C/O interfaces have smooth composition gradients in the low panel because of element diffusion. In the up panel, log($N^{2}$) is showed for the model. The $WD$ evolutions by the module make\_co\_wd in \texttt{MESA} are a little rough, as discussed above. There are discontinuities on the Brunt-V\"ais\"al\"a frequency for the core. Those discontinuities may cause false mode-trapping information in C/O core. There are two obvious spikes at log(1-$M_{\rm r}/M_{*}$) = -10 and -2 (the top abscissa) which correspond to the composition gradient region of H/He and He/C/O respectively. The composition gradient region forms a boundary of a standing wave and then a trapped mode emerges. Studying the mode trapping properties, a dimensionless mode inertia ($E$) was introduced by Christensen-Dalsgaard (2003) of
\begin{equation}
\ E = \frac{{4\pi\int_{0}^{R}[(|\tilde{\xi}_{r}(r)|^2+l(l+1)|\tilde{\xi}_{h}(r)|^2)]\rho_{0}r^2dr}}
{{M_{*}[|\tilde{\xi}_{r}(R)|^2+l(l+1)|\tilde{\xi}_{h}(R)|^2)]}}.
\end{equation}
\noindent In Eq.\,(5), $\tilde{\xi}_{r}(r)$ is the radial displacement, $\tilde{\xi}_{h}(r)$ is the horizontal displacement, and $\rho_{0}$ is the local density. The value of $E$ expresses kinetic energy normalized on the surface layer. The smaller the mode inertia of a mode, the larger the kinetic energy being confined on the surface. On the contrary, the larger the mode inertia of a mode, the larger the kinetic energy being confined on the core.

In Fig. 2, we show a diagram of period versus mode inertia, period versus period spacing, and kinetic energy distribution for $l$ = 2 modes. The H atmosphere for the model is log($M_{\rm H}/M_{*}$) = -10.0. The value of $k$ is from 22 to 45 and there are 24 modes between 800\,s and 1600\,s. We also calculate the $k$ = 46 mode which is larger than 1600\,s. The period spacing for the $k$ = 45 mode equals the period of $k$ = 46 minus the period of $k$ = 45. There are corresponding 24 period spacings showed in the middle panel. The three obvious minimal $E$s in the low panel basically correspond to the three groups of minimal period spacings in the middle panel. In the up panel, open dots, forks, and pluses represent the scale of kinetic energy distributed in H atmosphere, He layer, and C/O core respectively. The minimal mode inertias in the low panel correspond to the maximal open dots in the up panel. Basically, those modes ($P$($k$)) lead to minimal period spacings ($P$($k$+1)-$P$($k$) or $P$($k$)-$P$($k$-1)) in the middle panel. The maximal mode inertias in the low panel basically correspond to the maximal pluses in the up panel. In addition, the mode of $k$ = 37 with maximal mode inertia corresponds to a minimal period spacing. Namely, the minimal period spacings correspond to minimal or maximal $E$s. However, for the model of log($M_{\rm H}/M_{*}$) = -10.0, the modes with minimal $E$ dominate the minimal period spacings. The period spacings are from 22.45\,s to 44.73\,s, which are dispersive spanning more than 20\,s. Those minimal $E$s in the low panel correspond to the maximal open dots in the up panel. These modes do not have most or more than a half of kinetic energy distributed in H atmosphere. We suggest that they are partly, not wholly, trapped in H atmosphere. The standard deviation of the dispersion of period spacing $\sigma_{SD}$ is 5.37\,s for $l$ = 2 modes in the period range.

In Fig. 3, the model has H atmosphere of log($M_{\rm H}/M_{*}$) = -4. The value of $k$ is from 29 to 56 and there are 28 modes between 800\,s and 1600\,s. Basically, those maximal $E$s in the low panel correspond to the minimal period spacings in the middle panel, and maximal pluses in the up panel. The modes partly trapped in C/O core dominate the minimal period spacings. The H atmosphere is thick. Modes partly trapped in H layer are not obvious. The period spacings are from 23.99\,s to 32.48\,s, which spans less than 10\,s. Studying new chemical profiles for DAV stars by \texttt{LOCODE}, Althaus et al. (2010) revealed that the modes of minimal period spacings were associated with the modes partly trapped in the core regions. For our DAV star models, the phenomenon is obvious for thick H atmosphere. Though modes partly trapped in C/O core dominate the minimal period spacings, the period spacings are relatively concentrated. They are only spanning less than 10\,s. For the model of log($M_{\rm H}/M_{*}$) = -4.0, $\sigma_{SD}$ is 2.54\,s for $l$ = 2 modes.

In Fig. 4, we show $\sigma_{SD}$ values for models of log($M_{\rm H}/M_{*}$) from -10.0 to -4.0 with steps of 0.5. The values of $\sigma_{SD}$ are showed in pluses for $l$ = 1 modes and in forks for $l$ = 2 modes. The thinner the H atmosphere, basically, the larger the value of $\sigma_{SD}$. Thin H atmosphere makes obvious mode trapping effect with dispersive minimal period spacings and large values of $\sigma_{SD}$.

\subsection{The effect of different values of effective temperature and He layer mass}

\begin{figure}
\begin{center}
\includegraphics[width=8cm,angle=0]{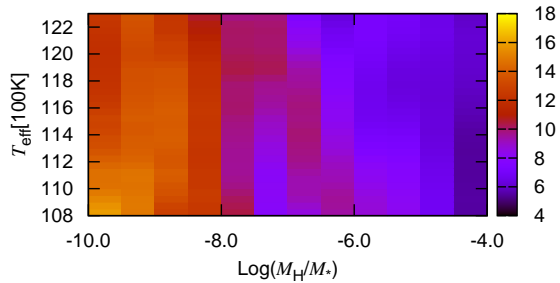}
\end{center}
\caption{Color diagram of $\sigma_{SD}$ for $l$ = 1 modes. The model parameters are $M_{*}$ = 0.600\,$M_{\odot}$, $T_{\rm eff}$ = 12300\,K to 10800\,K, log($M_{\rm He}/M_{*}$) = -2.0, and log($M_{\rm H}/M_{*}$) = -10.0 to -4.0.}
\end{figure}

\begin{figure}
\begin{center}
\includegraphics[width=8cm,angle=0]{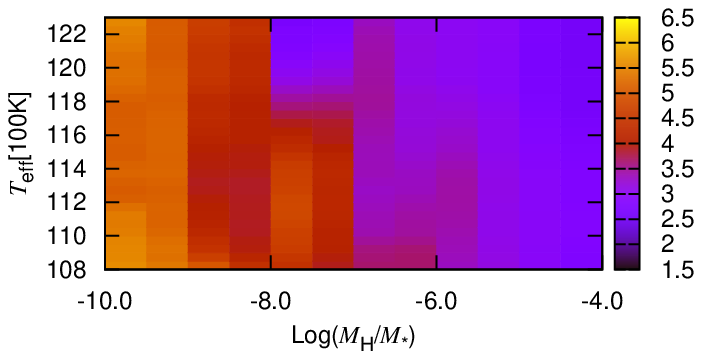}
\end{center}
\caption{Color diagram of $\sigma_{SD}$ for $l$ = 2 modes. The model parameters are $M_{*}$ = 0.600\,$M_{\odot}$, $T_{\rm eff}$ = 12300\,K to 10800\,K, log($M_{\rm He}/M_{*}$) = -2.0, and log($M_{\rm H}/M_{*}$) = -10.0 to -4.0.}
\end{figure}

\begin{figure}
\begin{center}
\includegraphics[width=8cm,angle=0]{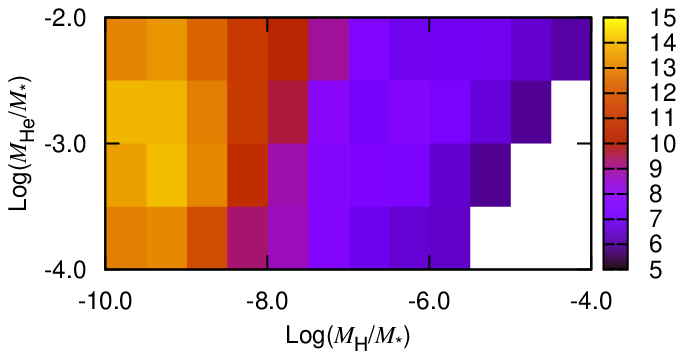}
\end{center}
\caption{Color diagram of $\sigma_{SD}$ for $l$ = 1 modes. The model parameters are $M_{*}$ = 0.600\,$M_{\odot}$, $T_{\rm eff}$ = 12000\,K, log($M_{\rm He}/M_{*}$) = -4.0 to -2.0, and log($M_{\rm H}/M_{*}$) = -10.0 to -4.0.}
\end{figure}

\begin{figure}
\begin{center}
\includegraphics[width=8cm,angle=0]{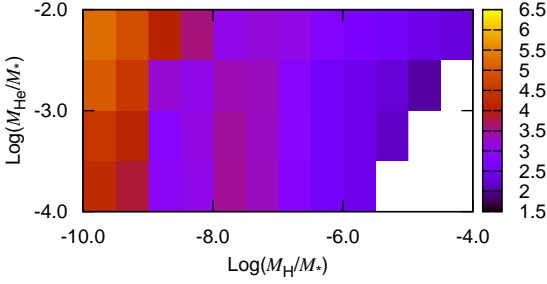}
\end{center}
\caption{Color diagram of $\sigma_{SD}$ for $l$ = 2 modes. The model parameters are $M_{*}$ = 0.600\,$M_{\odot}$, $T_{\rm eff}$ = 12000\,K, log($M_{\rm He}/M_{*}$) = -4.0 to -2.0, and log($M_{\rm H}/M_{*}$) = -10.0 to -4.0.}
\end{figure}

In Fig. 5 and 6, we show the dispersion of period spacing for $l$ = 1 and 2 modes respectively. The second group of DAV star models are used. The model parameters are $M_{*}$ = 0.600\,$M_{\odot}$, $T_{\rm eff}$ = 12300\,K to 10800\,K, log($M_{\rm He}/M_{*}$) = -2.0, and log($M_{\rm H}/M_{*}$) = -10.0 to -4.0. The parameter $\sigma_{SD}$ is basically from 4\,s to 18\,s for $l$ = 1 modes and 1.5\,s to 6.5\,s for $l$ = 2 modes. In Both figures, we can see that, from right to left, the color for $\sigma_{SD}$ changes from blue to orange. Namely, the thinner the H atmosphere, basically, the more dispersive the period spacing. However, from the top down, almost no change in color. The effective temperature has little influence on the dispersion of period spacing.

In Fig. 7 and 8, we show the dispersion of period spacing for $l$ = 1 and 2 modes respectively. The first group of DAV star models are used. The model parameters are $M_{*}$ = 0.600\,$M_{\odot}$, $T_{\rm eff}$ = 12000\,K, log($M_{\rm He}/M_{*}$) = -4.0 to -2.0, and log($M_{\rm H}/M_{*}$) = -10.0 to -4.0. The parameter $\sigma_{SD}$ is basically from 5\,s to 15\,s for $l$ = 1 modes and 1.5\,s to 6.5\,s for $l$ = 2 modes. The H atmosphere should be thinner than or equal to 1\% of the He layer in order to avoid overlapping transition regions (Arcoragi \& Fontaine 1980). Therefore, there are white blanks in Fig. 7 and 8. In general, the H atmosphere dominates the dispersion of period spacing. The He layer mass has little influence on the dispersion of period spacing.

\subsection{The effect of different values of total stellar mass}

\begin{figure}
\begin{center}
\includegraphics[width=8cm,angle=0]{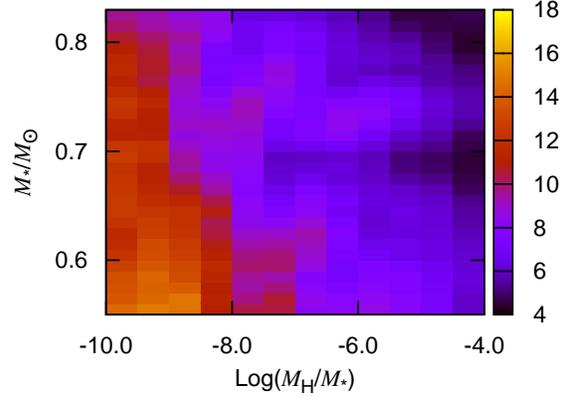}
\end{center}
\caption{Color diagram of $\sigma_{SD}$ for $l$ = 1 modes. The model parameters are $M_{*}$ = 0.550\,$M_{\odot}$ to 0.830\,$M_{\odot}$, $T_{\rm eff}$ = 12000\,K, log($M_{\rm He}/M_{*}$) = -2.0, and log($M_{\rm H}/M_{*}$) = -10.0 to -4.0.}
\end{figure}

\begin{figure}
\begin{center}
\includegraphics[width=8cm,angle=0]{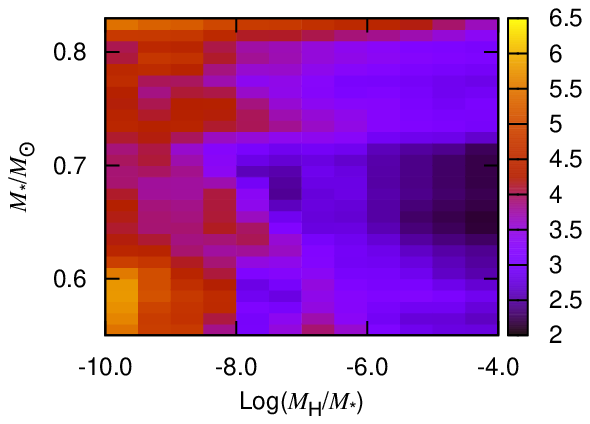}
\end{center}
\caption{Color diagram of $\sigma_{SD}$ for $l$ = 2 modes. The model parameters are $M_{*}$ = 0.550\,$M_{\odot}$ to 0.830\,$M_{\odot}$, $T_{\rm eff}$ = 12000\,K, log($M_{\rm He}/M_{*}$) = -2.0, and log($M_{\rm H}/M_{*}$) = -10.0 to -4.0.}
\end{figure}

\begin{figure}
\begin{center}
\includegraphics[width=8cm,angle=0]{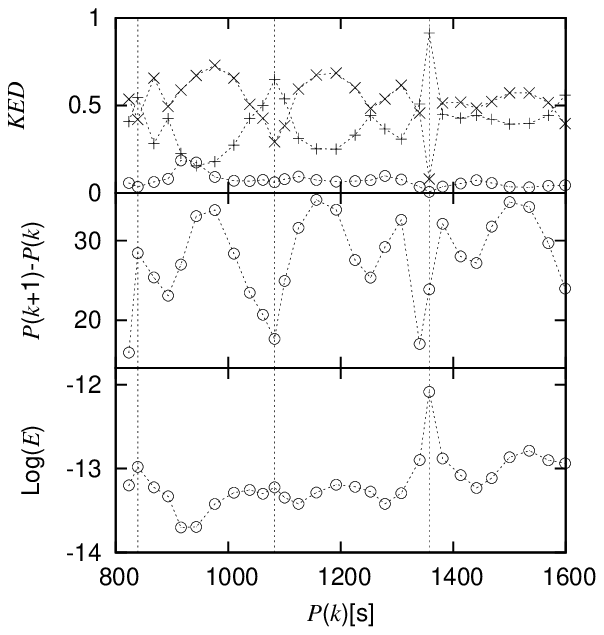}
\end{center}
\caption{Diagram of period versus mode inertia, period versus period spacing, and kinetic energy distribution ($KED$). The model parameters are $M_{*}$ = 0.830\,$M_{\odot}$, $T_{\rm eff}$ = 12000\,K, log($M_{\rm He}/M_{*}$) = -2.0, and log($M_{\rm H}/M_{*}$) = -10.0. The parameter $l$ equals 2 and $k$ is from 29 to 57. In the up panel, open dots, forks, and pluses represent the scale of kinetic energy distributed in H atmosphere, He layer, and C/O core respectively.}
\end{figure}

\begin{figure}
\begin{center}
\includegraphics[width=8cm,angle=0]{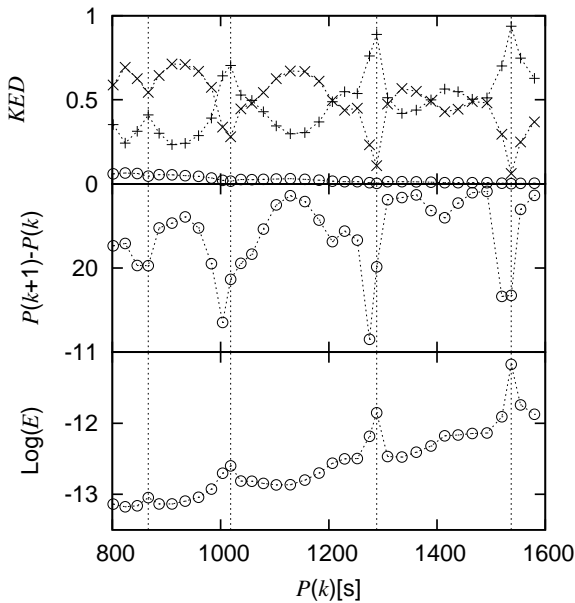}
\end{center}
\caption{Diagram of period versus mode inertia, period versus period spacing, and kinetic energy distribution ($KED$). The model parameters are $M_{*}$ = 0.830\,$M_{\odot}$, $T_{\rm eff}$ = 12000\,K, log($M_{\rm He}/M_{*}$) = -2.0, and log($M_{\rm H}/M_{*}$) = -4.0. The parameter $l$ equals 2 and $k$ is from 39 to 73. In the up panel, open dots, forks, and pluses represent the scale of kinetic energy distributed in H atmosphere, He layer, and C/O core respectively.}
\end{figure}

The effect of different values of total stellar mass is discussed in this subsection. The third group of DAV star models are used. With $T_{\rm eff}$ = 12000\,K, log($M_{\rm He}/M_{*}$) = -2.0, $M_{*}$ is from 0.550\,$M_{\odot}$ to 0.830\,$M_{\odot}$ and log($M_{\rm H}/M_{*}$) is from -10.0 to -4.0. For $l$ = 1 modes, $\sigma_{SD}$ is basically from 4\,s to 18\,s, as shown in Fig. 9. For $l$ = 2 modes, $\sigma_{SD}$ is basically from 2\,s to 6.5\,s, as shown in Fig. 10. In both figures, we can see that the thinner the H atmosphere, basically, the more dispersive the period spacing.

In Fig. 10, $\sigma_{SD}$ for $l$ = 2 modes seems almost large for models with $M_{*}$ $>$ 0.800\,$M_{\odot}$. For example, $\sigma_{SD}$ is from 5.64\,s to 3.57\,s when log($M_{\rm H}/M_{*}$) is from -10.0 to -4.0 for the model of $M_{*}$ = 0.830\,$M_{\odot}$. There are almost orange on the top of Fig. 10. This phenomenon is 'abnormal'. In Fig. 11 and 12, we show diagrams of period versus mode inertia, period versus period spacing, and kinetic energy distribution for $l$ = 2 modes. The model parameters are $M_{*}$ = 0.830\,$M_{\odot}$, log($M_{\rm H}/M_{*}$) = -10.0 and -4.0 for Fig. 11 and 12 respectively. Even log($M_{\rm H}/M_{*}$) equals -10.0, the mode trapping effect on H atmosphere is still not obvious. On the contrary, three modes partly trapped in C/O core dominate the minimal period spacings. In Fig. 12, for the model of $M_{*}$ = 0.830\,$M_{\odot}$ and log($M_{\rm H}/M_{*}$) = -4.0, four modes partly trapped in C/O core dominate the minimal period spacings. The modes partly trapped in H atmosphere do not dominate the minimal period spacings for DAV stars with $M_{*}$ basically larger than 0.800\,$M_{\odot}$. Therefore, $\sigma_{SD}$ show different schemes on the top of Fig. 10.

In Fig. 1, there are discontinuities on the core for Brunt-V\"ais\"al\"a frequency. These discontinuities are usually artificial because many physical processes will smooth them, such as over-shooting, under-shooting, thermohaline instabilities, diffusion, turbulence, and so on. Therefore, those modes who are partly trapped in C/O core for DAV stars with $M_{*}$ basically larger than 0.800\,$M_{\odot}$ are likely to be artificial. For observations, modes partly trapped in C/O core would have low probability to reach sufficiently high amplitude and would be absent in the light curves for a real star (Winget, Van Horn, \& Hansen 1981, Benvenuto et al. 2002). For the large mass DAV stars, the modes partly trapped in C/O core need further studies.

When $M_{*}$ is not as large as 0.800\,$M_{\odot}$, basically, modes partly trapped in H atmosphere dominate the minimal period spacings. Therefore, the thinner the H atmosphere, the more dispersive the period spacing. If some observed star have extremely large (small) dispersion of period spacing, it is likely to have corresponding thin (thick) H atmosphere. The parameter $\sigma_{SD}$ can be used to study the H atmosphere preliminarily. The obtained H atmosphere, especially for extreme values of $\sigma_{SD}$, can be compared with that obtained by detailed period to period fitting method (Castanheira, \& Kepler 2009, Romero et al. 2012).

\section{The average period spacing for DAV star models}

\begin{figure}
\begin{center}
\includegraphics[width=8cm,angle=0]{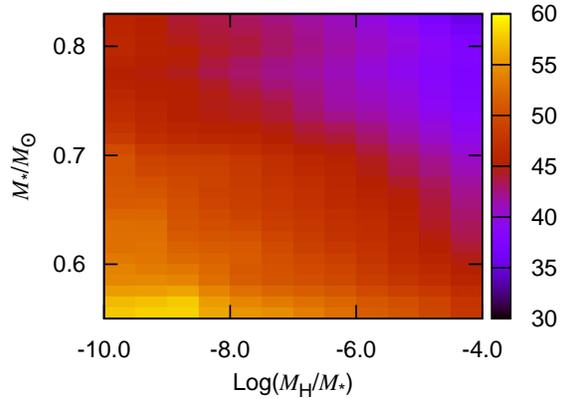}
\end{center}
\caption{Color diagram of $APS$ for $l$ = 1 modes. The model parameters are $M_{*}$ = 0.550\,$M_{\odot}$ to 0.830\,$M_{\odot}$, $T_{\rm eff}$ = 12000\,K, log($M_{\rm He}/M_{*}$) = -2.0, and log($M_{\rm H}/M_{*}$) = -10.0 to -4.0.}
\end{figure}

\begin{figure}
\begin{center}
\includegraphics[width=8cm,angle=0]{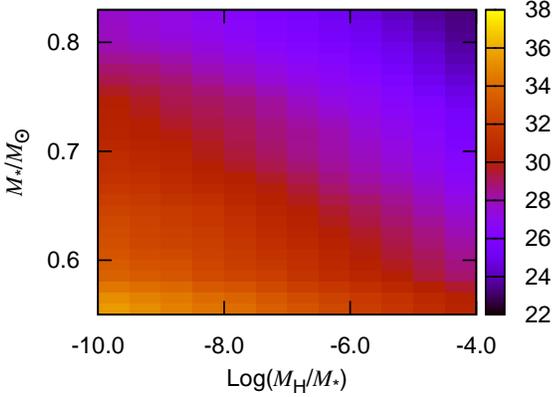}
\end{center}
\caption{Color diagram of $APS$ for $l$ = 2 modes. The model parameters are $M_{*}$ = 0.550\,$M_{\odot}$ to 0.830\,$M_{\odot}$, $T_{\rm eff}$ = 12000\,K, log($M_{\rm He}/M_{*}$) = -2.0, and log($M_{\rm H}/M_{*}$) = -10.0 to -4.0.}
\end{figure}

Studying the dispersion of period spacing, the average period spacings are also calculated for DAV star models. The average period spacing increases slowly with the effective temperature cooling down. It is a result of evolution. The thinner the H atmosphere or the He layer, the slightly larger the average period spacing. It is very obvious that the larger the total stellar mass, the smaller the average period spacing. In Fig. 13 and 14, we show the color diagram of $APS$ for $l$ = 1 and 2 modes respectively for the third group of models. The value of $APS$ is from 34.73\,s (right top in Fig. 13) to 58.49\,s (left bottom in Fig. 13) for $l$ = 1 modes. The value of $APS$ is from 23.00\,s (right top in Fig. 14) to 36.17\,s (left bottom in Fig. 14) for $l$ = 2 modes. For qualitative analysis, a large DAV star has small stellar radius and thus a large density. A large density corresponds to a large Brunt-V\"ais\"al\"a frequency and thus a small value of $APS$, according to Eq.\,(1). For $APS$ in Fig. 13 and 14, the total stellar mass and the H atmosphere are degenerate. The larger the total stellar mass, the smaller the value of $APS$. The thicker the H atmosphere, the smaller the value of $APS$. The dispersion of period spacing may remove the degeneration of $APS$ between total stellar mass and H atmosphere.

\section{The preliminary analysis on KUV03442+0719}

\begin{table}
\begin{center}
\begin{tabular}{lllllllllllll}
\hline
ID         $\,$        &$F$        &$\delta$$F$ &$m$        &Amplitude  &$P_{\rm rot}$ \\
           $\,$        &($\mu$Hz)  &($\mu$Hz)   &           &mmag       &(h)           \\
\hline
Quintuplets$\,$        &           &            &           &           &              \\
$f_{a\,-2}$$\,$(2010)  &685.93     &            &$\,$$-2$   &5.02       &              \\
           $\,$        &           &69.74       &           &           &6.76          \\
$f_{a\,0}$  $\,$ (2010)&755.67     &            &$\,$  $0$  &9.80       &              \\
           $\,$        &           &            &           &           &              \\
$f_{b\,-2}$$\,$(2011)  &632.43     &            &$\,$$-2$   &1.58       &              \\
           $\,$        &           &69.96       &           &           &6.75          \\
$f_{b\,0}$  $\,$ (2011)&702.39     &            &$\,$  $0$  &2.48       &              \\
           $\,$        &           &34.54       &           &           &6.64          \\
$f_{b\,+1}$$\,$(2011)  &736.93     &            &$\,$$+1$   &3.87       &              \\
           $\,$        &           &            &           &           &              \\
$f_{c\,-2}$$\,$(2011)  &642.54     &            &$\,$$-2$   &1.56       &              \\
           $\,$        &           &35.63       &           &           &6.75          \\
$f_{c\,-1}$$\,$(2011)  &678.17     &            &$\,$$-1$   &2.40       &              \\
           $\,$        &           &34.25       &           &           &6.46          \\
$f_{c\,0}$  $\,$ (2011)&712.42     &            &$\,$  $0$  &3.39       &              \\
           $\,$        &           &33.66       &           &           &6.81          \\
$f_{c\,+1}$$\,$(2011)  &746.08     &            &$\,$$+1$   &1.71       &              \\
           $\,$        &           &34.04       &           &           &6.71          \\
$f_{c\,+2}$$\,$(2011)  &780.12     &            &$\,$$+2$   &6.72       &              \\
           $\,$        &           &            &           &           &              \\
$f_{d\,-1}$$\,$(2011)  &743.32     &            &$\,$$-1$   &2.35       &              \\
           $\,$        &           &33.68       &           &           &6.93          \\
$f_{d\,0}$  $\,$ (2011)&777.00     &            &$\,$  $0$  &5.53       &              \\
           $\,$        &           &            &           &           &              \\
$f_{e\,-2}$$\,$(2011)  &729.95     &            &$\,$$-2$   &2.51       &              \\
           $\,$        &           &69.83       &           &           &6.74          \\
$f_{e\,0}$  $\,$ (2011)&799.78     &            &$\,$  $0$  &4.95       &              \\
           $\,$        &           &34.69       &           &           &6.62          \\
$f_{e\,+1}$$\,$(2011)  &834.47     &            &$\,$$+1$   &2.71       &              \\
           $\,$        &           &            &           &           &              \\
$f_{f\,0}$  $\,$ (2011)&856.84     &            &$\,$  $0$  &2.58       &              \\
           $\,$        &           &68.47       &           &           &6.65          \\
$f_{f\,+2}$$\,$(2011)  &925.31     &            &$\,$$+2$   &2.10       &              \\
           $\,$        &           &            &           &           &              \\
Triplets   $\,$        &           &            &           &           &              \\
           $\,$        &           &            &           &           &              \\
$f_{g\,-1}$$\,$(2010)  &1003.07    &            &$\,$$-1$   &3.92       &              \\
           $\,$        &           &21.49       &           &           &6.46          \\
$f_{g\,0}$  $\,$ (2010)&1024.56    &            &$\,$  $0$  &4.74       &              \\
           $\,$        &           &            &           &           &              \\
$f_{h\,0}$  $\,$ (2012)&843.44     &            &$\,$  $0$  &2.60       &              \\
           $\,$        &           &21.10       &           &           &6.60          \\
$f_{h\,+1}$$\,$(2012)  &864.54     &            &$\,$$+1$   &2.54       &              \\
\hline
\end{tabular}
\caption{Possible rotational splits for KUV03442+0719 identified by Su et al. (2014b). In the header, $F$ is frequency and $\delta$$F$ is frequency splitting value.}
\end{center}
\end{table}

In Table 2, we show the possible rotational splits for a DAV star named KUV03442+0719 identified by Su et al. (2014b). Both Pesnell (1985) and Brassard, Fontaine, \& Wesemael (1995) reported the amplitude relations among components of triplets and quintuplets based on the inclination angle of rotation axis. According to the relations, a set of complete quintuplets, five sets of incomplete quintuplets, and two sets of incomplete triplets were identified by Su et al. (2014b) in Table 2. According to the amplitudes for modes of $f_{c\,-2}$, $f_{c\,-1}$, $f_{c\, 0}$, $f_{c\,+1}$, $f_{c\,+2}$ and $f_{e\,-2}$, $f_{e\, 0}$, $f_{e\,+1}$, the center components of KUV 03442+0719 may have relative large amplitudes, which correspond to a small inclination angle of the rotation axis. Su et al. (2014b) hypothesized that the $m$ = 0 components had relative large amplitudes. If there was a singlet mode observed, it would be identified as a $m$ = 0 component. In addition, the amplitude of $f_{c\,+2}$ shows a large value, which is not agree with the relations for small inclination angle reported by Pesnell (1985). Both Handler et al. (2008) and Fu et al. (2013) reported the phenomenon studying DAV star EC14012-1446 and HS0507+0434B respectively. The amplitude ratio might be not only restricted by the geometrical effect, as Fu et al. (2013) explained.

\begin{table*}
\begin{center}
\begin{tabular}{lcccccccccccccc}
\hline
ID         &$l$($k$) &$F$       &$P$      &$\delta$$P$($\delta$$k$)  &ID        &$l$($k$)     &$F$       &$P$      &$\delta$$P$($\delta$$k$)   \\
           &         &($\mu$Hz) &(s)      &(s)                       &          &             &($\mu$Hz) &(s)      &(s)                        \\
\hline
                        &2(68)    &400.83    &2494.82  &             &                        &1(43)        &410.00    &2439.02  &             \\
                        &         &          &         &69.16(4)     &                        &             &          &         &632.24(19)   \\
                        &2(64)    &412.26    &2425.66  &             &                        &1(24)        &553.47    &1806.78  &             \\
                        &         &          &         &916.89(47)   &                        &             &          &         &120.27(3)    \\
                        &2(17)    &662.79    &1508.77  &             &                        &1(21)        &592.94    &1686.51  &             \\
                        &         &          &         &85.07(4)     &                        &             &          &         &290.46(9)    \\
$f_{b\,0}$   $\,$ (2011)&2(13)    &702.39    &1423.70  &             &                        &1(12)        &716.31    &1396.05  &             \\
                        &         &          &         &20.02(1)     &                        &             &          &         &69.29(2)     \\
$f_{c\,0}$   $\,$ (2011)&2(12)    &712.42    &1403.68  &             &                        &1(10)        &753.71    &1326.76  &             \\
                        &         &          &         &54.46(3)     &                        &             &          &         &37.16(1)     \\
                        &2(9)     &741.17    &1349.22  &             &                        &1(9)         &775.43    &1289.60  &             \\
                        &         &          &         &25.90(1)     &                        &             &          &         &52.17(1)     \\
$f_{a\,0}$   $\,$ (2010)&2(8)     &755.67    &1323.32  &             &                        &1(8)         &808.13    &1237.43  &             \\
                        &         &          &         &36.32(2)     &                        &             &          &         &51.80(2)     \\
$f_{d\,0}$   $\,$ (2011)&2(6)     &777.00    &1287.00  &             &$f_{h\,0}$   $\,$ (2010)&1(6)         &843.44    &1185.63  &             \\
                        &         &          &         &36.65(2)     &                        &             &          &         &46.03(1)     \\
$f_{e\,0}$   $\,$ (2011)&2(4)     &799.78    &1250.35  &             &                        &1(5)         &877.50    &1139.60  &             \\
                        &         &          &         &57.76(3)     &                        &             &          &         &115.71(4)    \\
                        &2(1)     &838.51    &1192.59  &             &                        &1(1)         &976.67    &1023.89  &             \\
                        &         &          &         &25.52(1)     &                        &             &          &         &37.86(1)     \\
$f_{f\,0}$   $\,$ (2011)&2(0)     &856.84    &1167.07  &             &$f_{g\,0}$   $\,$ (2012)&1(0)         &1024.56   &976.03   &             \\
                        &         &          &         &30.58(2)     &                        &             &          &         &67.42(2)     \\
                        &2(-2)    &879.90    &1136.49  &             &                        &1(-2)        &1100.58   &908.61   &             \\
                        &         &          &         &38.99(2)     &                        &             &          &         &476.82(14)   \\
                        &2(-4)    &911.16    &1097.50  &             &                        &1(-16)       &2315.97   &431.79   &             \\
                        &         &          &         &66.45(3)     &                        &             &          &         &             \\
                        &2(-7)    &969.79    &1031.15  &             &                        &             &          &         &             \\
                        &         &          &         &58.25(3)     &                        &             &          &         &             \\
                        &2(-10)   &1027.85   &972.90   &             &                        &             &          &         &             \\
                        &         &          &         &544.24(28)   &                        &             &          &         &             \\
                        &2(-38)   &2332.85   &428.66   &             &                        &             &          &         &             \\
\hline
\end{tabular}
\caption{Results of mode identifications by Su et al. (2014b). In the header, $k$ is relative radial order, $P$ is period, $\delta$$P$ is period difference value, and $\delta$$k$ is difference of corresponding relative radial orders.}
\end{center}
\end{table*}

According to the frequency splitting relations, the asymptotic period spacing for $g$-modes, and the period spacing relations for $l$ = 1 and 2 modes, Su et al. (2014b) made detailed mode identifications for KUV03442+0719, including the relative $k$ values. Their identified results are showed in Table 3. In addition, we calculate the period difference value ($\delta$$p$) between two identified adjacent modes in Table 3. The corresponding difference ($\delta$$k$) between relative radial orders is calculated. Most of the identified modes for KUV03442+0719 are in the range of 800\,s to 1600\,s. The values of $k$ reflect relative relations between different radial orders, not real radial orders. The average period spacing was derived as 19.54\,s for $l$ = 2 modes and 34.07\,s for $l$ = 1 modes by least square fittings. For the calculated values of $APS$ in Fig. 13 and 14, the smallest is 23.00\,s for $l$ = 2 modes and 34.73\,s for $l$ = 1 modes. For $l$ = 2 modes, 19.54\,s is obviously smaller than 23.00\,s. For $l$ = 1 modes, 34.07\,s is close to 34.73\,s. The total stellar mass for KUV03442+0719 is likely to be around the order of 0.8\,$M_{\odot}$. One possible scenario is that there should be more modes partly trapped in C/O core, which can reduce the $APS$ values for models. The core compositions are very important and need further in-depth studies for asteroseismological work on DAV stars. The other possible scenario is that the average period spacing for KUV03442+0719 is not so small essentially. When more modes are observed through future observations, some mode identifications may be revised and then the $APS$ values may increase.

The mode of $l$ = 1, $k$ = 8 was observed in 2011 (1237.43\,s) and 2012 (1238.00\,s) by Su et al. (2014b). The mode of $l$ = 1, $k$ = 0 was observed in 2010 (976.03\,s) and 2012 (976.61\,s). We take the modes with high amplitude and S/N to calculate the period differences in Table 3. In the period range from 800\,s to 1600\,s, a parameter $\sigma_{SD}'$ is calculated by
\begin{equation}
\sigma_{SD}'=\sqrt{\frac{1}{n} \sum_{\delta k}(\delta k)(\frac{\delta P}{\delta k}-APS)^2}.
\end{equation}
\noindent In Eq.\,(6), $n$ is the sum of $\delta k$. In the period range from 800\,s to 1600\,s, $n$ is 27 ($k$=17-$k$=-10) for $l$ = 2 modes and 14 ($k$=12-$k$=-2) for $l$ = 1 modes. The operation of sum should be calculated $n$ times. However, there are not $n$+1 periods observed in the period range. In Table 3, $f_{a\,0}$ and $f_{d\,0}$ were identified as $l$ = 2, $k$ (relative radial order) = 8 and 6 respectively. For those two modes, $\delta P$/$\delta k$ = 36.32/2\,s is used twice for the sum calculations. Mathematically, $\sigma_{SD}'$ is less than or equal to $\sigma_{SD}$. For the DAV star KUV03442+0719, those missing modes are not likely to be equally spaced. Therefore, $\sigma_{SD}'$ is less than $\sigma_{SD}$. According to Eq\.(3), in the period range from 800\,s to 1600\,s, $APS$ is (1508.77\,s-972.90\,s)/27=19.847\,s for $l$ = 2 modes and (1396.05\,s-908.61\,s)/14=34.817\,s for $l$ = 1 modes. After careful calculations, $\sigma_{SD}'$ is 2.40\,s for $l$ = 2 modes and 7.28\,s for $l$ = 1 modes. The values of $\sigma_{SD}'$ show that the H atmosphere of KUV03442+0719 is not very thick. Comparing the values of $\sigma_{SD}'$ for $l$ = 1 and 2 modes to the values of $\sigma_{SD}$ from Fig. 5 to Fig. 10, log($M_{\rm H}/M_{*}$) is less than or equal to -5.5 for KUV03442+0719.

In addition, we also calculate a parameter $\sigma_{SD}''$ by
\begin{equation}
\sigma_{SD}''=\sqrt{\frac{1}{n''} \sum_{n''}(\frac{\delta P}{\delta k}-APS)^2}.
\end{equation}
\noindent In Eq.\,(7), $n''$ is the number of period differences. In the period range from 800\,s to 1600\,s, $n''$ is 12 for $l$ = 2 modes and 8 for $l$ = 1 modes. For the modes of $f_{a\,0}$ and $f_{d\,0}$, $\delta P$/$\delta k$ = 36.32/2\,s is used once for the sum calculations. One period difference is calculated once. The parameter $\sigma_{SD}''$ is used to study the dispersion of period spacing for KUV03442+0719 approximately. In the period range of 800\,s to 1600\,s, $APS$ is also 19.847\,s for $l$ = 2 modes and 34.817\,s for $l$ = 1 modes. After careful calculations, $\sigma_{SD}''$ is 2.97\,s for $l$ = 2 modes and 8.34\,s for $l$ = 1 modes. In Fig. 4, the linear relationship between H atmosphere and standard deviation is poor. Comparing the values of $\sigma_{SD}''$ for $l$ = 1 and 2 modes to the values of $\sigma_{SD}$ from Fig. 5 to Fig. 10, log($M_{\rm H}/M_{*}$) is basically from -8.5 to -5.5 for KUV03442+0719. The H atmosphere can be compared with that obtained by detailed period to period fitting method for KUV03442+0719 in the future work.

In Table 3, $APS$ is 19.847\,s for $l$ = 2 modes and 34.817\,s for $l$ = 1 modes in the period range from 800\,s to 1600\,s. The average period spacing for all the modes in Table 3 was 19.54\,s for $l$ = 2 modes and 34.07\,s for $l$ = 1 modes derived by Su et al. (2014b). However, there is no real observed period spacing less than 19.54\,s for $l$ = 2 modes and less than 34.07\,s for $l$ = 1 modes. We only observe $\delta$$P$ values less than corresponding $\delta$$k$ times of 19.54\,s for $l$ = 2 modes and 34.07\,s for $l$ = 1 modes. Namely, some missing modes dominate the minimal period spacings. Modes partly trapped in H atmosphere are easily observed and modes partly trapped in C/O core are not easily observed. Those missing modes who dominate the minimal period spacings may happen to be partly trapped in C/O core. KUV03442+0719 may be the first star to 'observe' modes partly trapped in C/O core, according to the mode identifications by Su et al. (2014b).

\section{Discussion and conclusions}

DAV stars are $g$-mode pulsators. The $g$-modes follow an asymptotic period spacing law. Trapped or partly trapped modes insert themselves into the equally spaced periods and then lead to small period spacings. Therefore, those trapped or partly trapped modes should be closely related to the dispersion of period spacing. Studying the dispersion of period spacing may be helpful to study mode trapping properties and inner structures of DAV stars. Based on these ideas, we evolve three groups of DAV star models by \texttt{WDEC} taking the element diffusion effect into account. The $WD$ cores are directly from $WD$ models evolved by \texttt{MESA}, which are results of thermonuclear burning. Those DAV star models have historically viable core composition profiles. In theory, the core compositions are results of thermonuclear burning. Taking the core profiles from fully evolutionary ($MS$ stars to $WD$ stars) $WD$s is an attempt direction. However, it is a preliminary attempt to insert core profiles from $WD$ models evolved by \texttt{MESA} into \texttt{WDEC} to evolve $WD$ models. Some flaws need to be improved in the future work, such as the discontinuities on the Brunt-V\"ais\"al\"a frequency for the core, especially for large mass DAV stars.

Studying the dispersion of period spacing on the evolved DAV star models, we find that the H atmosphere dominates the dispersion of period spacing. The thinner the H atmosphere, the more dispersive the period spacing. Both the effective temperature and the He layer mass have little influence on the dispersion of period spacing. In addition, the average period spacing decreases with increasing the total stellar mass and the H atmosphere mass. For some observed star, the average period spacing can be used to constrain the stellar mass preliminary. If a very large dispersion of period spacing is observed, a very thin H atmosphere will be derived. However, in our calculations, modes partly trapped in C/O core dominate the dispersion of period spacing for large mass DAV stars. Those modes should be difficult to observe. The phenomenon is not unique. It was also reported by Brassard, Fontaine, \& Wesemael (1995), Benvenuto et al. (2002), Althaus et al. (2010). Therefore, the core compositions are very important for asteroseismological work on DAV stars. Modes partly trapped in C/O core may dominate the dispersion of period spacing.

KUV03442+0719, as a DAV star, was observed in 2010, 2011, and 2012 by Su et al. (2014b). A set of complete quintuplets, five sets of incomplete quintuplets, and two sets of incomplete triplets were identified. Mode identifications are relatively reliable based on the frequency splitting relations. Su et al. (2014b) derived an average period spacing of 19.54\,s for $l$ = 2 modes and 34.07\,s for $l$ = 1 modes. We calculate the period differences between adjacent modes for $l$ = 2 and 1. The small average period spacings for observed modes indicate that KUV03442+0719 should be a large mass DAV star, such as the order of 0.8\,$M_{\odot}$. The parameters $\sigma_{SD}'$ and $\sigma_{SD}''$ are used to study the dispersion of period spacing for KUV03442+0719. If $\sigma_{SD}'$ is large enough, the H atmosphere should be very thin. However, $\sigma_{SD}'$ is not large for KUV03442+0719. The values of $\sigma_{SD}'$ indicate that log($M_{\rm H}/M_{*}$) is less than or equal to -5.5. The parameter $\sigma_{SD}''$ is 2.97\,s for $l$ = 2 modes and 8.34\,s for $l$ = 1 modes. The values show that KUV03442+0719 has a H atmosphere of log($M_{\rm H}/M_{*}$) = -8.5 to -5.5. In addition, for the values of $\delta$$P$ with $\delta$$k$ in Table 3, the trapped modes are missing in the scheme of period difference. There is no period difference smaller than the average period spacing for $l$ = 1 and 2 modes respectively. The missing modes make small period differences which are smaller than corresponding $\delta$$k$ times of the $APS$ values. Therefore, KUV03442+0719 may be the first star to 'observe' modes partly trapped in C/O core.

\section{Acknowledgements}

This work is supported by the National Natural Science Foundation (11563011) and the Yunnan Applied Basic Research Project Foundation (2015FD044). We are very grateful to X. J. Lai, T. Wu, J. J. Guo, J. Su, Q. S. Zhang, and Y. Li for their kindly discussion and suggestions.

\label{lastpage}


\begin{thebibliography}{99}
\bibitem[\protect\citeauthoryear{Althaus}{2010}]{b1} Althaus L. G., C$\acute{o}$rsico A. H., Bischoff-Kim A., et al., 2010, ApJ, 717, 897
\bibitem[\protect\citeauthoryear{Angulo}{1999}]{b1} Angulo C., Arnould M., Rayet M., et al., 1999, Nucl. Phys. A, 656, 3
\bibitem[\protect\citeauthoryear{Arcoragi}{1980}]{b1} Arcoragi J. P., Fontaine G., 1980, ApJ, 242, 1208
\bibitem[\protect\citeauthoryear{Benvenuto}{2002}]{b1} Benvenuto O. G., C$\acute{o}$rsico A. H., Althaus L. G., Serenelli A. M., 2002, MNRAS, 335, 480
\bibitem[\protect\citeauthoryear{Bergeron}{1995}]{b1} Bergeron P., Wesemael F., Lamontagne R., et al., 1995, ApJ, 449, 258
\bibitem[\protect\citeauthoryear{Brassard}{1991}]{b1} Brassard P., Fontaine G., Wesemael F., et al., 1991, ApJ, 367, 601
\bibitem[\protect\citeauthoryear{Brassard}{1992a}]{b1} Brassard P., Fontaine G., Wesemael F., Hansen C. J., 1992a, ApJS, 80, 369
\bibitem[\protect\citeauthoryear{Brassard}{1992b}]{b1} Brassard P., Fontaine G., Wesemael F., Tassoul M., 1992b, ApJS, 81, 747
\bibitem[\protect\citeauthoryear{Brassard}{1995}]{b1} Brassard P., Fontaine G., Wesemael F., et al., 1995, ApJ, 96, 545
\bibitem[\protect\citeauthoryear{Bischoff-Kim}{2011}]{b1} Bischoff-Kim A., Metcalfe T. S., 2011, MNRAS, 414, 404
\bibitem[\protect\citeauthoryear{Brickhill}{1975}]{b1} Brickhill A. F., 1975, MNRAS, 170, 405
\bibitem[\protect\citeauthoryear{Castanheira}{2010}]{b1} Castanheira B. G., Kepler S. O., Kleinman S.J., Nitta A., Fraga L., 2010, MNRAS, 405, 2561
\bibitem[\protect\citeauthoryear{Castanheira}{2009}]{b1} Castanheira B. G., Kepler S. O., 2009, MNRAS, 396, 1709
\bibitem[\protect\citeauthoryear{Caughlan}{1988}]{b1} Caughlan G. R., Fowler W. A., 1988, At. Data Nucl. Data Tables, 40, 283
\bibitem[\protect\citeauthoryear{Chen}{2014}]{b1} Chen Y. H., Li Y., 2014, MNRAS, 443, 3477
\bibitem[\protect\citeauthoryear{Christensen-Dalsgaard}{2003}]{b1} Christensen-Dalsgaard J., Lecture Notes on Stellar Oscillations, Fifth Edition, May 2003, http://astro.phys.au.dk/~jcd/oscilnotes/
\bibitem[\protect\citeauthoryear{Costa}{2008}]{b1} Costa J. E. S., Kepler S. O., Winget D. E., et al., 2008, A\&A, 477, 627
\bibitem[\protect\citeauthoryear{Cox}{1989}]{b1} Cox A. N., Guzik J. A., Kidman R. B., 1989, ApJ, 342, 1187
\bibitem[\protect\citeauthoryear{C$\acute{o}$rsico}{2002}]{b1} C$\acute{o}$rsico A. H., Althaus L. G., Benvenuto O. G., Serenelli A. M., 2002, A\&A, 387, 531
\bibitem[\protect\citeauthoryear{Fontaine}{2001}]{b1} Fontaine G., Brassard P., Bergeron P., 2001, PASP, 113, 409
\bibitem[\protect\citeauthoryear{Fu}{2013}]{b1} Fu J.-N., Dolez N., Vauclair G., et al., 2013, MNRAS, 429, 1585
\bibitem[\protect\citeauthoryear{Gianninas}{2005}]{b1} Gianninas A., Bergeron P., Fontaine G., 2005, ApJ, 631, 1100
\bibitem[\protect\citeauthoryear{Gianninas}{2006}]{b1} Gianninas A., Bergeron P., Fontaine G., 2006, ApJ, 132, 831
\bibitem[\protect\citeauthoryear{Gianninas}{2011}]{b1} Gianninas A., Bergeron P., Ruiz M. T., 2011, ApJ, 743, 138
\bibitem[\protect\citeauthoryear{Handler}{2008}]{b1} Handler G., Romero-Colmenero E., Provencal J. L., et al., 2008, MNRAS, 388, 1444
\bibitem[\protect\citeauthoryear{Kanaan}{2005}]{b1} Kanaan A., Nitta A., Winget D. E., 2005, A\&A, 432, 219
\bibitem[\protect\citeauthoryear{Kepler}{2000}]{b1} Kepler S. O., Reed M. D., Kawaler S. D., Bradley P. A., 2000, ApJ, 534, L185
\bibitem[\protect\citeauthoryear{Kutter}{1969}]{b1} Kutter, G. S., Savedoff, M. P., 1969, ApJ, 156, 1021
\bibitem[\protect\citeauthoryear{Lamb}{1975}]{b1} Lamb D. Q., Van Horn H. M., 1975, ApJ, 200, 306
\bibitem[\protect\citeauthoryear{Li}{1992a}]{b1} Li Y., 1992a, A\&A, 257, 133
\bibitem[\protect\citeauthoryear{Li}{1992b}]{b1} Li Y., 1992b, A\&A, 257, 145
\bibitem[\protect\citeauthoryear{Montgomery}{1999}]{b1} Montgomery M. H., Winget D. E., 1999, ApJ, 526, 976
\bibitem[\protect\citeauthoryear{Paquette}{1986}]{b1} Paquette C., Pelletier C., Fontaine G., Michaud G., 1986a, ApJS, 61, 177
\bibitem[\protect\citeauthoryear{Paquette}{1986}]{b1} Paquette C., Pelletier C., Fontaine G., Michaud G., 1986b, ApJS, 61, 197
\bibitem[\protect\citeauthoryear{Paxton}{2011}]{b1} Paxton B., Bildsten L., Dotter A., et al., 2011, ApJS, 192, 3
\bibitem[\protect\citeauthoryear{Pesnell}{1985}]{b1} Pesnell W. D., 1985, ApJ, 292, 238
\bibitem[\protect\citeauthoryear{Romero}{2012}]{b1} Romero A. D., C$\acute{o}$rsico A. H., Althaus L. G., et al., 2012, MNRAS, 420, 1462
\bibitem[\protect\citeauthoryear{Su}{2014a}]{b1} Su J., Li Y., Fu J. -N., Li, C., 2014a, MNRAS, 437, 2566
\bibitem[\protect\citeauthoryear{Su}{2014b}]{b1} Su J., Li Y., Fu J. -N., 2014b, New Astronomy, 33, 52
\bibitem[\protect\citeauthoryear{Tassoul}{1980}]{b1} Tassoul M., 1980, ApJS, 43, 469
\bibitem[\protect\citeauthoryear{Tassoul}{1990}]{b1} Tassoul M., Fontaine G., Winget D. E., 1990, ApJS, 72, 335
\bibitem[\protect\citeauthoryear{Thoul}{1994}]{b1} Thoul A. A., Bahcall J. N., Loeb A., 1994, ApJ, 421, 828
\bibitem[\protect\citeauthoryear{Winget}{1981}]{b1} Winget D. E., Van Horn H. M., Hansen C. J., 1981, ApJ, 245, L33
\bibitem[\protect\citeauthoryear{Winget}{2008}]{b1} Winget D. E., Kepler S. O., 2008, ARAA, 46, 157
\bibitem[\protect\citeauthoryear{Wood}{1990}]{b1} Wood M. A., 1990, Astero-archaeology: Reading the Galactic History Recorded in the White Dwarf Stars,
Ph.D. Thesis, Texas Univ., Austin
\bibitem[\protect\citeauthoryear{Zhang}{2008}]{b1} Zhang Q. S., May 2008, Application of the turbulent convection model to the sun and the problem of density inversion (in Chinese), Master thesis, Yunnan Obsertaries, Chinese academy of Sciences.
\end{thebibliography}
\end{document}